\documentclass{iopjournal}

\usepackage[utf8]{inputenc}
\usepackage[english]{babel}
\usepackage{amsmath}
\usepackage{amsfonts}
\usepackage{amssymb}
\usepackage{amsthm}
\usepackage{bm}
\usepackage{float}
\usepackage{algorithm}
\usepackage{algpseudocode}
\usepackage{ragged2e}
\AtBeginDocument{\justifying}
\def\l{\langle}
\def\r{\rangle}
\def\mb{\bar{m}}

\def\qb{\bar{q}}

\newcommand{\SOMMA}[2]{\sum\limits_{#1}^{#2}}

\newtheorem{remark}{Remark}

\allowdisplaybreaks

\begin{document}

\articletype{Paper} 

\title{Partial annealing and pattern decorrelation in associative neural networks}

\author{Linda Albanese$^{1,2,*}$\orcid{0000-0003-1783-0775}, Andrea Alessandrelli$^{1,2}$\orcid{0000-0003-4583-1417}, Adriano Barra$^{3,4, 5}$\orcid{0000-0003-4255-7678}, Silvio Franz$^{1,2,6}$\orcid{0000-0001-8300-8443} and Federico Ricci-Tersenghi$^{4,7,8}$\orcid{0000-0003-4970-7376}}

\affil{$^1$Dipartimento di Matematica e Fisica, Università del Salento, 73100 Lecce, Italy}

\affil{$^2$Istituto Nazionale di Fisica Nucleare, Sezione di Lecce, 73100 Lecce, Italy}

\affil{$^3$Dipartimento di Scienze di Base e Applicazioni all'Ingegneria, Sapienza Università di Roma, 00161 Rome, Italy}

\affil{$^4$Istituto Nazionale di Fisica Nucleare, Sezione di Roma 1, 00185 Rome, Italy}

\affil{$^5$CNR-Nanotec, unità di Lecce, 73100 Lecce, Italy}

\affil{$^6$Université Paris-Saclay, CNRS, LPTMS, 91405 Orsay, France}

\affil{$^7$Dipartimento di Fisica, Sapienza Università di Roma, 00185 Rome, Italy}

\affil{$^8$CNR-Nanotec, unità di Roma, 00185 Rome, Italy}

\affil{$^*$Author to whom any correspondence should be addressed.}

\email{linda.albanese@unisalento.it}

\keywords{Partial annealing, Neural networks, Guerra interpolation, Statistical mechanics}

\begin{abstract}
\justifying
Using the Hopfield model as a benchmark case, the present work focuses on the investigation of partially annealed associative neural networks, wherein neural dynamics is coupled to slowly evolving patterns within the two-temperature-two-timescale framework. This setting inherently introduces a real parameter $n$, reminiscent of the number of replicas in the celebrated replica trick, that tunes the separation of timescales and the effective interaction between fast (i.e. the neurons) and slow (i.e. the synapses) degrees of freedom.  By adapting Guerra’s interpolation to the case, we derive the free energy without relying on analytical continuation. The obtained results  demonstrate that negative values of $n$ induce a progressive decorrelation of the stored patterns, thereby effectively reducing interference, promoting orthogonal configurations and ultimately conferring to the network the maximal storage $\alpha_c =1$. \\
Numerical simulations based on a mean field Monte Carlo dynamics have been employed to confirm this scenario and prove that partial annealing restores retrieval in challenging regimes, such as in the presence of biased patterns, outperforming standard decorrelation methods.
These findings underscore the notion of partial annealing as an adaptive mechanism for enhancing memory organisation and retrieval in complex systems.
\end{abstract}

\section{Introduction}
The statistical mechanics of complex systems has long constituted an effective framework for the quantitative analysis of associative memory networks, in particular for studying pattern recognition, memory storage, and information processing in neural networks \cite{MPV, hertz1991introduction, AGS, HopKro1, Coolen}.
Disordered systems characteristically manifest intricate energy landscapes, exhibiting a multitude of metastable configurations and frustration \cite{MPV, nishimori2001statistical, franz2000ultrametricity}. In statistical mechanics, a fundamental conceptual distinction is made in the description of such systems between \textit{quenched} and \textit{annealed} \textit{disorder}. In the quenched scenario, disorder variables evolve on timescales that are much longer than those associated with the dynamical degrees of freedom. As a result, on the pratical level, they can be regarded as effectively frozen. Conversely, in annealed systems, both sets of variables equilibrate on comparable timescales. \\ While these two limiting cases are analytically tractable, many real systems operate in an intermediate regime, in which disorder evolves slowly but cannot be considered as completely frozen
\cite{Pierluse,Tonno}. In these regards, despite the theoretical analysis of these systems via the \textit{replica method} was carried on already a long time ago \cite{dotsenko1994partial}, some approaches  such as the Guerra interpolation \cite{guerra_broken, mezard2003cavity, barra2010replica,Albanese2021, agliari2022nonlinear}, which have played a pivotal role in determining the thermodynamic properties of mean field spin glass models, were lacking for \textit{partial annealing} and are developed in this paper. 

Moreover, the theoretical analysis of these systems is founded on a range of analytical techniques, including the \textit{replica method} and interpolation schemes such as the Guerra interpolation \cite{guerra_broken, mezard2003cavity, barra2010replica,Albanese2021, agliari2022nonlinear}, which have played a pivotal role in determining the thermodynamic properties of mean field spin glass models. 

\par\medskip
The concept of \textit{partial annealing} \cite{dotsenko1994partial} was introduced to capture this intermediate regime with precision. In such systems, slow variables, which are often associated with interactions or couplings, are not fixed but evolve according to an effective thermodynamic principle. This is typically characterized by a temperature distinct from that one governing the fast dynamical variables. This approach was originally proposed as a means to study disordered systems in which the couplings themselves are capable of adapting or reassessing, giving rise to effects such as overfrustration and non-trivial feedback between slow and fast degrees of freedom (see \cite{dotsenko1994partial}). The incorporation of two interacting thermal processes within the framework of partial annealing naturally interpolates between the quenched and annealed limits.

\par\medskip
In the realm of theoretical physics, the statistical mechanics of partially annealed systems can be formulated within the two temperatures framework. In this setting, the fast degrees of freedom (i.e. the neurons) are considered to reach equilibrium at a given temperature, whereas the slow variables (i.e. the synapses) are assumed to evolve at a separate (colder) temperature, hence their dynamics is expected to be slower with respect to the neural one \cite{DavidTon,Tonno}. The effective free energy governing the slow variables is thus shown to involve the partition function of the fast subsystem, raised to a power determined by the ratio between the two temperatures. This configuration lends the replica method particular analytical efficacy, as the temperature ratio can be interpreted as an effective replica number, thereby facilitating the application of standard techniques developed for the study of disordered systems. In accordance with the established framework, the analysis is conventionally conducted under the Replica Symmetric (RS) assumption, which entails the treatment of all replicas as equivalent in the saddle point solution. However, in many disordered systems, this symmetry may become unstable \cite{de1978stability}, requiring the more general Replica Symmetry Breaking (RSB) scheme \cite{MPV, AABO-JPA2020}, which captures the hierarchical organization of metastable states in the free energy landscape.

This formalism was soon applied to the study of adaptive and neural systems, in which synaptic interactions may evolve on timescales that are longer than those characterizing neuronal dynamics. In the early stages of research, analyses were conducted on neural networks with partially annealed couplings. These analyzes demonstrated that the adaptation of interactions has the capacity to cause substantial alterations in both the structure of the attractors and the phase diagram of the system \cite{feldman1994partially}. Later investigations examined these effects in more specific neural network models, emphasising how the interaction between slow synaptic evolution and fast neural activity can give rise to non-trivial forms of collective behaviour \cite{uezu2010statistical}.

\par\medskip
In recent years related concepts have subsequently emerged in machine learning, particularly in the study of Hopfield networks, Boltzmann machines, and ensemble learning, where learning processes have been analysed using tools from the statistical mechanics of disordered systems, stochastic stability, and replica theory \cite{Contucci,Tulinski2026}. In particular, the training dynamics and hyper-parameter optimisation of models such as Restricted Boltzmann Machines can be interpreted within frameworks that resemble partially annealed systems. In such frameworks, parameters and dynamical variables evolve at different effective temperatures or timescales  \cite{agliari2018rbm, fachechi2025fundamental}. This viewpoint establishes a connection between classical disordered systems and contemporary learning theory, providing new perspectives on glassy behaviour, optimisation landscapes, and the interpretability of trained models.

\par\medskip
The objective of this study is to provide a theoretical description of this intermediate regime by means of Guerra's interpolation scheme, within both the RS 
frameworks. 
This study builds on the foundational research of Dotsenko et al \cite{dotsenko1994partial}, investigating the role of the number of replicas in this context, which is naturally extended to take real values.

\par\medskip
The manuscript is organized as follows. After a general introduction of the Hopfield model in Sec. \ref{sec:generalities}, we apply Guerra's interpolating on the replica-dependent free energy expression for RS approximation in Sec. \ref{sec:RS}. Then, we focus on some numerical results in Sec. \ref{sec:numerical} to demonstrate that negative values of the parameter $n$ promote a gradual reduction in the correlations among the stored patterns. Some mathematical appendices regarding some computations 
close the manuscript. 

\section{Generalities}
\label{sec:generalities}

Let $\{\bm\xi^{\mu}\}_{\mu=1}^P$ be a collection of $P$ patterns of length $N$, with i.i.d.\ Rademacher entries, namely
\begin{equation}
\mathbb {P}(\xi_{i}^{\mu})
=\frac{1}{2}\delta(\xi_{i}^{\mu}-1)+\frac{1}{2}\delta(\xi_{i}^{\mu}+1),
\qquad \forall\,i=1,\dots,N,\quad \mu=1,\dots,P.
\label{eq:patterns_general}
\end{equation}
These random variables are used to build the Hebbian interaction matrix
\begin{equation}
J_{ij}(\bm \xi)
=\frac{1}{N}\sum_{\mu=1}^{P}\xi_{i}^\mu\xi_{j}^\mu,
\qquad i,j=1,\dots,N,
\label{eq:Jij}
\end{equation}
which encodes the pairwise correlations among the stored patterns and models the effective interactions among neurons in the network.

The neuronal configuration is denoted by
\begin{equation}
\bm{\sigma} = (\sigma_1,\dots,\sigma_N)\in \Omega,
\qquad \Omega=\{-1,+1\}^{N},
\end{equation}
and the model is governed by the Hamiltonian
\begin{align}
H_{N}(\bm{\sigma}\,|\,\bm{\xi})
:=
-\frac{1}{2}\sum_{i,j=1}^{N}
J_{ij}(\bm \xi)\,\sigma_{i}\sigma_{j}.
\label{eq:hamiltonian}
\end{align}
The retrieval quality of pattern $\bm\xi^\mu$ is measured by the corresponding \textit{Mattis magnetization}
\begin{equation}
m_{\mu}(\bm{\sigma} \vert \bm \xi)
:=\frac{1}{N}\sum_{i=1}^{N}\xi_i^\mu \sigma_i,
\qquad \mu=1,\dots,P.
\label{eq:Mattis}
\end{equation}
Using the definition of the Hebbian matrix \eqref{eq:Jij}, the Hamiltonian can be rewritten in terms of these order parameters \eqref{eq:Mattis} as
\begin{equation}
H_{N}(\bm{\sigma}\,|\,\bm{\xi})
=
-\frac{N}{2}\sum_{\mu=1}^{P}\big(m_\mu(\bm{\sigma} \vert \bm \xi)\big)^2.
\label{eq:hamiltonian_magn}
\end{equation}
Allowing for stochastic noise, controlled by the inverse temperature $\beta=T^{-1}\in\mathbb{R}_+$, the configuration space $\Omega$ is endowed with the (random) Boltzmann-Gibbs measure
\begin{equation}
\mathbb{P}_{N}(\bm{\sigma}|\bm{\xi})
:=\frac{\exp\big(-\beta\,H_{N}(\bm{\sigma}\,|\,\bm{\xi})\big)}
{Z_{N}(\beta\,|\,\bm{\xi})},
\label{BGmeasure}
\end{equation}
where
\begin{equation}
Z_{N}(\beta\,|\,\bm{\xi})
:=\sum_{\bm{\sigma}\in\Omega}
\exp\big[-\beta\,H_{N}(\bm{\sigma}\,|\,\bm{\xi})\big]
\label{partition-function}
\end{equation}
is the partition function. Accordingly, for any observable $g(\bm\sigma)$, the Boltzmann-Gibbs average is defined by
\begin{equation}
\omega_{N,\bm\xi}(g(\bm\sigma))
:=
\frac{1}{Z_{N}(\beta\,|\,\bm{\xi})}
\sum_{\bm{\sigma}\in\Omega} g(\bm\sigma)\,
\exp\big(-\beta\,H_{N}(\bm{\sigma}\,|\,\bm{\xi})\big),
\end{equation}
and the full quenched average is
\begin{equation}
\label{eq:totalaver}
\langle g(\boldsymbol \sigma) \rangle
=
\mathbb{E}_{\bm \xi}\Big[\omega_{N,\bm\xi}(g(\bm\sigma))\Big].
\end{equation}

Throughout the paper we work in the high-storage regime, namely we assume that the load
\begin{equation}
\alpha := \lim_{N\to\infty}\frac{P}{N} > 0
\label{eq:load}
\end{equation}
remains finite in the thermodynamic limit.

\begin{remark}[Slow vs.\ fast degrees of freedom in associative neural networks]
In associative neural networks with coupled neuron-synapse dynamics, it is useful to distinguish between two well-separated timescales:
\begin{itemize}
    \item the \emph{fast degrees of freedom} are the neuronal states $\{\sigma_i\}$, which relax quickly toward a local or quasi-equilibrium configuration for a given realization of the synaptic couplings;
    \item the \emph{slow degrees of freedom} are the synaptic variables $\{J_{ij}\}$, or equivalently the stored patterns that parameterize them, whose evolution takes place on a much longer timescale.
\end{itemize}
This separation implies that, in the long run, the synaptic dynamics is effectively driven by macroscopic observables of the fast subsystem, such as the free energy or the neuron-neuron correlations computed after the neurons have relaxed. Therefore, the fast variables perform retrieval dynamics for almost frozen couplings, whereas the slow variables reshape the energy landscape itself.
\end{remark}

The above viewpoint naturally suggests a generalization of the standard Hopfield setting. Instead of treating the patterns $\bm\xi$ as quenched disorder from the outset, one may regard them as slow dynamical variables and ask how their evolution is influenced by the thermodynamics of the fast neuronal subsystem. This is precisely the perspective underlying the \textit{partial annealing} picture proposed in \cite{dotsenko1994partial}, and it is also the starting point of the present work.

\par\medskip
More specifically, our goal is to develop this idea in two complementary directions. On the one hand, we formulate the theory directly for a real parameter $n$, without restricting the analysis to integer replicas and without relying on analytic continuation arguments (see also \cite{B-war4}) and explore the network's behaviour also for negative values of $n$. On the other hand, in the next section we will complement the analytical picture with a numerical algorithm aimed at probing the slow rearrangement of the patterns, thereby testing the physical scenario suggested in \cite{dotsenko1994partial}, namely that negative values of $n$ induce a tendency toward pattern decorrelation and progressive orthogonalization.

To formalize this two timescales picture, we start from the Hamiltonian of the Hopfield model \eqref{eq:hamiltonian} and its partition function \eqref{partition-function}. For a fixed realization of the patterns $\bm\xi$, the corresponding free energy is
\begin{equation}
    f_{N,P}(\beta|\bm\xi):=-\frac{1}{\beta}\log Z_N(\beta|\bm\xi).
    \label{eq:free_beta}
\end{equation}
If the patterns are no longer averaged out as quenched disorder, but are instead promoted to slow dynamical variables, then the neuronal free energy at fixed $\bm\xi$ naturally plays the role of an \textit{effective energy} for the pattern sector. Denoting by $\Omega'=\{-1,+1\}^{N\times P}$
the pattern configuration space, we define
\begin{equation}
    \mathcal{E}_{N,P}(\bm\xi)=f_{N,P}(\beta|\bm\xi),
\end{equation}
and introduce a second statistical level in which $\bm\xi$ is sampled according to the Boltzmann weight
\begin{equation}
    \mathbb{P}_{N,P}'(\bm\xi)
    =
    \frac{\exp\big(-\beta'\mathcal{E}_{N,P}(\bm\xi)\big)}
    {\mathcal{Z}_{N,P}(\beta')},
\end{equation}
where
\begin{equation}
    \mathcal{Z}_{N,P}(\beta')
    :=
    \mathbb{E}\left[\exp\left(-\beta'\mathcal{E}_{N,P}(\bm\xi)\right)\right]
    \label{eq:part_beta1}
\end{equation}
is the pattern level partition function, $\beta'$ is a second inverse temperature, in general different from $\beta$, and $\mathbb{E}[\cdot]$ denotes expectation with respect to the \textit{a priori} law of the patterns.

Using \eqref{eq:free_beta}, we can rewrite \eqref{eq:part_beta1} as
\begin{equation}
    \mathcal{Z}_{N,P}(\beta')
    =
    \mathbb{E}\left[
    \exp\left(\frac{\beta'}{\beta}\log Z_N(\beta|\bm\xi)\right)
    \right]
    =
    \mathbb{E}\Big[Z_N^n(\beta|\bm\xi)\Big],
    \label{eq:part_beta1_new}
\end{equation}
where we have set
\begin{equation}
    n=\frac{\beta'}{\beta}.
\end{equation}
Accordingly, it is natural to introduce the associated intensive free energy
\begin{equation}
    f_{N,n}(\beta)
    :=
    -\frac{1}{N\beta'}\log \mathcal{Z}_{N,P}(\beta')
    =
    -\frac{1}{N\beta'}\log\mathbb{E}\Big[Z_N^n(\beta|\bm\xi)\Big].
\end{equation}

\begin{remark}
Interpreting
\[
n=\frac{\beta'}{\beta}=\frac{T}{T'},
\]
the parameter $n$ measures the relative thermal scale of the slow and fast sectors, and therefore quantifies the interplay between pattern dynamics and neuronal equilibration. In particular, two limiting cases are immediately recovered:
\begin{itemize}
    \item \emph{Quenched regime}: $n\to 0$, corresponding to fixed $T$ and $T'\to\infty$ (equivalently, $\beta'\to 0$). In this limit the pattern dynamics becomes indifferent to the neuronal thermodynamics, and one recovers the usual quenched setting;
    \item \emph{Annealed regime}: $n\to 1$, corresponding to $T'=T$ (equivalently, $\beta'=\beta$). In this case the patterns evolve on the same thermal scale as the neurons, yielding the fully annealed regime.
\end{itemize}
\end{remark}

This representation shows that the natural object arising from the two levels construction is $\mathbb{E}_{\bm \xi}Z_N^n$, and hence the associated free energy
\begin{equation}
    f_{n,N}(\beta)
    =
    -\frac{1}{\beta n N}\log \mathbb{E}_{\bm\xi} Z_N^n(\beta \vert \bm\xi).
    \label{eq:An_finite_size}
\end{equation}
Its thermodynamic limit
\begin{equation}
    f_n(\beta)
    =
    -\lim_{N\to\infty}f_{n,N}(\beta)
    \label{eq:An}
\end{equation}
will be referred to as the \emph{pseudo-replicated free energy}. The terminology is meant to emphasize that, although the same expression appears in replica computations when $n\in\mathbb N$, here $n$ is regarded from the outset as a fixed real parameter with a direct physical interpretation in terms of timescale separation.

\begin{remark}
When $n$ is a positive integer, \eqref{eq:An} coincides with the replicated free energy arising in the standard replica method \cite{parisi2000origin}. In that framework, one exploits the fact that the moments $\mathbb{E}_{\bm\xi}Z_N^n(\beta|\bm\xi)$ are often easier to handle than the averaged logarithm $\mathbb{E}_{\bm\xi}\log Z_N(\beta|\bm\xi)$, and then formally reconstructs the quenched free energy through the limit
\begin{equation}
    -\lim_{N\to\infty}\frac{1}{\beta N}\mathbb{E}_{\bm\xi}\log Z_N(\beta \vert \bm\xi)
    =
    \lim_{n\to 0} f_n(\beta).
\end{equation}
The standard procedure therefore requires computing the moments for integer $n$, extending the resulting expression to real $n$ by analytic continuation, and finally taking the limit $n\to 0$. In the present work, by contrast, no such continuation is needed: $n$ is treated directly as a real control parameter, and the resulting expression for $f_n(\beta)$ will be derived by means of Guerra's interpolation \cite{guerra_broken}. Moreover, the case $n=1$ consistently reproduces the annealed regime. While the case $n=0$ capture the quenched one as expected. Moreover, recent rigorous progress on negative replicas in Gaussian spin-glass models has clarified convexity and differentiability properties of logarithmic moment generating functions $\frac{1}{n N}\log \mathbb{E} Z_N^n$, providing quantitative results related to the Dotsenko-Franz-Mézard analysis in the Sherrington-Kirkpatrick model \cite{chen2026gaussian}. Although obtained in a different setting, these results further motivate the
study of the negative-replica sector as a meaningful thermodynamic regime.
\end{remark}

From the physical viewpoint, the most intriguing regime is $n<0$, corresponding to a negative effective temperature in the slow sector. As already suggested in \cite{dotsenko1994partial}, this regime biases the pattern distribution toward over-frustrated configurations and is expected to induce a non-trivial reorganization of the stored patterns. One of the motivations of the present work is precisely to clarify this scenario analytically and to test it numerically in the finite-size system.

\section{Interpolating replicas - RS assumption}
\label{sec:RS}
We now turn to the derivation of $f_n(\beta)$ by Guerra's interpolation method. Since our interest lies in the retrieval regime, where at least one pattern --without loss of generality $\bm\xi^1$-- is selected as the candidate retrieved pattern, we separate its contribution from the remaining $P-1$ patterns, which act as a source of slow noise. Accordingly, the partition function can be rewritten as
\begin{equation}
    Z_N(\beta \vert \bm \xi)
    =
    \sum_{\bm{\sigma}\in\Omega}
    \exp\Bigg[
    \frac{\beta}{2N}\sum_{i,j =1}^{N}\xi_i^1\xi_j^1 \sigma_i\sigma_j
    +
    \frac{\beta}{2N}\sum_{\mu>1}^{P}\sum_{i,j =1}^{N}\xi_i^\mu\xi_j^\mu \sigma_i\sigma_j
    \Bigg].
\end{equation}
The first contribution is the \textit{signal term} associated with the Mattis magnetization of the retrieved pattern, while the second one collects the noise generated by all the non-retrieved patterns.

In order to implement Guerra's interpolation scheme, we next apply the \textit{Hubbard-Stratonovich transformation} to the noise term and obtain
\begin{equation}
    Z_N(\beta \vert \bm \xi)
    =
    \sum_{\bm{\sigma}\in\Omega}\int d\mu(z)\,
    \exp\Bigg[
    \frac{\beta}{2N}\sum_{i,j =1}^{N}\xi_i^1\xi_j^1 \sigma_i\sigma_j
    +
    \sqrt{\frac{\beta}{N}}\sum_{\mu>1}^{P}\sum_{i =1}^{N}\xi_i^\mu \sigma_i z_\mu
    \Bigg],
    \label{eq:hubbard}
\end{equation}
where
\[
d\mu(\bm z)= \prod_{\mu>1}^P\frac{dz_\mu}{\sqrt{2\pi}}e^{-z_\mu^2/2}.
\]
This transformation can be interpreted as a mapping of the original slow-noise contribution of the Hebbian network into a Restricted Boltzmann Machine, where the $N$ visible spins $\sigma_i$ interact with $P-1$ Gaussian hidden variables $z_\mu$ through the couplings $\xi_i^\mu$ \cite{B-war1}.

\begin{remark}
\label{rem:universality}
In the following we shall exploit the universality property of quenched noise in spin glasses \cite{CarmonaWu, Genovese}: independently of whether the pattern entries are digital (e.g.\ Boolean) or analog (e.g.\ Gaussian), as long as the distribution is centered, symmetric, and has finite variance, their contribution to the quenched pressure is asymptotically the same in the thermodynamic limit. This property is specific to the high-storage regime and is not guaranteed when $\lim_{N\to\infty}P/N=0$.
\end{remark}

Therefore, in view of Remark \ref{rem:universality}, from now on we assume that the candidate retrieved pattern $\bm\xi^1$ is Rademacher, whereas the remaining patterns $\bm\xi^\mu$, $\mu=2,\dots,P$, are Gaussian. This mixed formulation is technically convenient and leaves the thermodynamic picture unchanged in the thermodynamic limit.

\par\medskip
To describe the macroscopic behavior of the model, we introduce the standard order parameters: the Mattis magnetization \eqref{eq:Mattis} with respect to $\bm\xi^1$, which quantifies retrieval, and the \textit{two-replicas overlap}
\begin{equation}
    q_{ab}(\bm \sigma) = \frac{1}{N} \sum_{i=1}^N \sigma_i^a \sigma_i^b,
    \label{eq:order_q}
\end{equation}
which measures the level of slow noise experienced by the network. Since the Hubbard-Stratonovich transformation introduces the auxiliary variables $\bm z=\{z_\mu\}_{\mu=2,\dots,P}$, we also define the corresponding \textit{hidden-layer overlap}
\begin{equation}
    p_{ab}(\bm z)= \frac{1}{P-1}\sum_{\mu>1}^{P}z_\mu^{a}z_\mu^{b},
    \label{eq:order_p}
\end{equation}
which captures the interference induced by the non-retrieved patterns\footnote{Throughout the present analysis, the number of auxiliary patterns $z_\mu$ corresponding to the slow noise will be $P-1$. However, since our primary interest lies in the thermodynamic limit of the model, and this quantity will invariably appear in the form of a ratio with $N$, we shall not distinguish between the ratios $P/N$ and $(P-1)/N$. Accordingly, where convenient, we shall write $P/N$ in place of $(P-1)/N$.}. In what follows, the order parameters are understood to depend on $\bm{\sigma}$, $\bm{\xi}$ and $\bm{z}$.

\par\medskip
Our goal is now to derive the expression of the pseudo-replicated free energy \eqref{eq:An} by means of Guerra's interpolation method.

Following \cite{guerra_broken}, we introduce an interpolating parameter $t\in[0,1]$ and define the interpolating partition function $Z_N(\beta\vert\bm\xi;t)$, explicitly given in \eqref{eq:interpolating_Z}. We then consider the associated finite size interpolating pseudo-replicated free energy
\begin{equation}
\label{eq:interpol_pressure}
    f_{N,n}(\beta \vert t)
    =
    -\frac{1}{\beta n N}\log \mathbb{E}_{\bm\xi} Z_N^n(\beta|\bm\xi, \bm J, \bm Y;t),
\end{equation}
where the interpolating partition function is defined as
\begin{equation}
\begin{aligned}
    Z_N(\beta\vert \bm\xi, \bm J, \bm Y;t)
    &=
    \sum_{\bm\sigma\in\Omega}\int d\mu(z_\mu)\,
    \exp\Bigg[-\beta H_N(\bm\sigma,\bm z \vert \bm\xi, \bm J, \bm Y;t)\Bigg]
    \\
    &=
    \sum_{\bm\sigma\in\Omega}\int d\mu(z_\mu)\,
    \exp\Bigg[
    t \frac{\beta N}{2}(m_1)^2
    +(1-t)\psi N m_1
    +\sqrt{t}\sqrt{\frac{\beta}{N}}\sum_{\mu,i}\xi_i^\mu \sigma_i z_\mu
    \\
    &\hspace{3.5cm}
    +\sqrt{1-t}\,B\sum_{\mu}Y_{\mu} z_{\mu}
    +\sqrt{1-t}\,A\sum_i J_i \sigma_i
    +\frac{1-t}{2}C\sum_\mu z_\mu^2
    \Bigg],
\end{aligned}
\label{eq:interpolating_Z}
\end{equation}
with $t\in[0,1]$, the constants $\psi$, $A$, $B$, and $C$ will be fixed \textit{a posteriori}, and $\bm Y=\{Y_\mu\}_{\mu=1,\dots,P}$ and $\bm J=\{J_i\}_{i=1,\dots,N}$ are i.i.d.\ standard Gaussian random variables. The corresponding interpolating Hamiltonian reads
\begin{equation}
\begin{aligned}
    H_N(\bm\sigma,\bm z \vert \bm\xi, \bm J, \bm Y;t)
    &=
    -t\frac{N}{2}(m_1)^2
    -\frac{(1-t)\psi}{\beta}N m_1
    -\sqrt{t}\sqrt{\frac{1}{N\beta}}\sum_{\mu,i}\xi_i^\mu \sigma_i z_\mu
    \\
    &\qquad
    -\frac{\sqrt{1-t}\,B}{\beta}\sum_\mu Y_\mu z_\mu
    -\frac{\sqrt{1-t}\,A}{\beta}\sum_i J_i \sigma_i
    -\frac{1-t}{2\beta}C\sum_\mu z_\mu^2.
\end{aligned}
\end{equation}

At $t=1$ we recover the original model, namely $f_{N,n}(\beta \vert t=1)=f_{N,n}(\beta)$. In $t=0$, instead, the system reduces to a one-body model, which is explicitly solvable: neurons decouple and interact only with effective external fields chosen to reproduce, at the level of the relevant low-order statistics, the internal fields generated by the entire interacting system.

The relation between the two endpoints is provided by the Fundamental Theorem of Calculus (FTC):
\begin{equation}
    \label{eq:T_o_C}
    f_{N,n}(t=1)
    =
    f_{N,n}(t=0)+\int_0^1 ds\,\frac{\partial f_{N,n}(t)}{\partial t}\Big|_{t=s}.
\end{equation}
We now analyze the two contributions separately.

\paragraph{One-body contribution.}
The first term is readily computed. Setting $t=0$ in \eqref{eq:interpolating_Z}, we obtain
\begin{equation}
\mathbb{E}Z_N^n(\beta \vert \bm J, \bm Y;t=0)
=
\mathbb{E}\Bigg[
\sum_{\bm\sigma\in\Omega}\int d\mu(\bm z)\,
\exp\Bigg(
\psi N m_1
+B\sum_\mu Y_\mu z_\mu
+A\sum_i J_i \sigma_i
+\frac{C}{2}\sum_\mu z_\mu^2
\Bigg)
\Bigg]^n.
\end{equation}
At this stage the contributions depending on $\bm\sigma$ and on $\bm z$ factorize. By performing the Gaussian integrations and using standard algebraic manipulations, one finds
\begin{equation}
\mathbb{E}Z_N^n(\beta \vert \bm\xi, \bm J, \bm Y;t=0)
=
2^{nN}(1-C)^{-nP/2}
\prod_{i=1}^N
\mathbb{E}\Big[\cosh\big(\psi \xi^1 + AJ\big)\Big]^n
\exp\Bigg\{
P\log\Bigg[\frac{1}{\sqrt{1-\frac{nB^2}{1-C}}}\Bigg]
\Bigg\},
\end{equation}
and therefore
\begin{equation}
\begin{aligned}
   f_{n}(\beta; t=0)
   &=
   -\frac{1}{\beta}\log 2
   -\frac{1}{\beta n}\log\mathbb{E}_{x}\cosh^n\Big(\psi+x\sqrt{A^2}\Big)
   \\
   &\qquad
   +\frac{P}{2N\beta}\log(1-C)
   -\frac{P}{2N\beta n}\log\Bigg(1-\frac{nB^2}{1-C}\Bigg).
\end{aligned}
\label{eq:one_body}
\end{equation}
The details of this computation are reported in Appendix~\ref{app:tderiv}.

\paragraph{Streaming term.}
At finite volume, the derivative with respect to $t$ of the interpolating pseudo-replicated free energy is
\begin{equation}
\begin{aligned}
    \partial_t f_{n,N}(\beta; t)
    =
    \frac{1}{N}\mathbb{E}\Bigg[
    W_{n,N}(\beta \vert \bm\xi, \bm J, \bm Y;t)\,
    \omega\Big(\partial_t H_N(\bm\sigma,\bm z \vert \bm\xi, \bm J, \bm Y;t)\Big)
    \Bigg],
\end{aligned}
\label{eq:dtf_0}
\end{equation}
where
\begin{align}
W_{n,N}(\beta \vert \bm\xi, \bm J, \bm Y;t)
:=
\frac{Z_N^n(\beta\vert \bm\xi, \bm J, \bm Y;t)}
{\mathbb{E} Z_N^n(\beta\vert \bm\xi, \bm J, \bm Y;t)}
\end{align}
and
\begin{equation}
    \l \cdot \r := \mathbb{E}\Bigg[
    W_{n,N}(\beta \vert \bm\xi, \bm J, \bm Y;t)\,
    \omega\Big(\cdot\Big)
    \Bigg].
\end{equation}
Moreover,
\begin{equation}
\begin{aligned}
   \partial_t H_N(\bm\sigma,\bm z|\bm\xi, \bm J, \bm Y;t)
   &=
   -\frac{N}{2}(m_1)^2
   +\frac{\psi}{\beta}Nm_1
   -\frac{1}{2}\sqrt{\frac{1}{tN\beta}}\sum_{\mu,i}\xi_i^\mu \sigma_i z_\mu
   \\
   &\qquad
   +\frac{B}{2\beta\sqrt{1-t}}\sum_\mu Y_{\mu} z_{\mu}
   +\frac{A}{2\beta\sqrt{1-t}}\sum_i J_i \sigma_i
   +\frac{1}{2\beta}C\sum_\mu z_\mu^2.
\end{aligned}
\label{eq:dtf_1}
\end{equation}
As shown in Appendix~\ref{app:tderiv}, this expression can be rewritten as
\begin{equation}
\begin{aligned}
   \partial_t f_{n,N}(\beta; t)
   &=
   - \frac{1}{2}\Bigg[
   \langle (m_1)^2\rangle
   -\frac{2\psi}{\beta}\langle m_1\rangle
   \Bigg]
   -\frac{P}{2N}\Big[
   \langle p_{11}\rangle
   -(1-n)\langle q_{12}p_{12}\rangle
   \Big]
   \\
   &\qquad
   +\frac{B^2P}{2\beta N}\Big[
   \langle p_{11}\rangle-(1-n)\langle p_{12}\rangle
   \Big]
   +\frac{A^2}{2\beta}\Big[
   1-(1-n)\langle q_{12}\rangle
   \Big]
   +\frac{P}{2\beta N}C\langle p_{11}\rangle.
\end{aligned}
\label{eq:dtf}
\end{equation}

We now impose the RS \textit{ansatz}, namely that the relevant order parameters self-average around deterministic values in the thermodynamic limit. More precisely, for any order parameter $X$ with limiting value $\bar X$, we assume
\begin{equation}
    \lim_{N\to\infty}\mathbb{P}_N(X)=\delta(X-\bar X).
\end{equation}
Under this assumption, the thermodynamic limit of the streaming term becomes
\begin{equation}
\begin{aligned}
    \lim_{N\to\infty}\partial_t f_{n,N}(\beta; t)
    =
    \partial_t f_{n,\alpha}(\beta; t)
    =
    \frac{1}{2}\bar m^2
    +\frac{\alpha}{2}\bar p(1-\bar q)
    +\frac{\alpha}{2}n\bar p\,\bar q,
\end{aligned}
\label{eq:dtff}
\end{equation}
provided that the interpolating parameters are chosen as
\begin{equation}
    \begin{aligned}
         \psi = \beta \bar m, \qquad
         C = \beta(1-\bar q), \qquad
         B^2 = \beta \bar q, \qquad
         A^2 = \alpha\beta \bar p,
    \end{aligned}
\end{equation}
and we have used the definition of the load in \eqref{eq:load}.

\paragraph{RS pseudo-replicated free energy and self-consistency equations.}
Substituting \eqref{eq:one_body} and \eqref{eq:dtff} into \eqref{eq:T_o_C}, we obtain the RS expression of the pseudo-replicated free energy in the thermodynamic limit:
\begin{equation}
\label{eq:free_n}
\begin{aligned}
     f_{n, \alpha}(\beta)
     &=
     -\frac{1}{\beta}\log 2
     -\frac{1}{\beta n}\log\mathbb{E}_{x}\cosh^n\Big(
     \beta \bar m + x\sqrt{\beta \alpha \bar p}
     \Big)
     +\frac{\alpha}{2\beta}\dfrac{n-1}{n}\log\Big(1-\beta(1-\bar q)\Big)
     \\
     &\qquad
     +\frac{\alpha}{2\beta n}\log\Big(
     1-\beta(1-\bar q)-n\beta \bar q
     \Big)
     +\frac{1}{2}\bar m^2
     +\frac{\alpha}{2}\bar p(1-\bar q)
     +\frac{\alpha}{2}n\bar p\,\bar q
\end{aligned}
\end{equation}
The corresponding order parameters satisfy the self-consistency equations
\begin{equation}
\label{eq:self_n}
    \begin{aligned}
        \bar{m}
        &=
        \frac{
        \mathbb{E}_{x}\Big\{
        \cosh^n\big(
        \beta \bar{m} + x\sqrt{\beta\alpha\bar{p}}
        \big)
        \tanh\big(
        \beta \bar{m} + x\sqrt{\beta\alpha\bar{p}}
        \big)
        \Big\}
        }{
        \mathbb{E}_{x}\Big\{
        \cosh^n\big(
        \beta \bar{m} + x\sqrt{\beta\alpha\bar{p}}
        \big)
        \Big\}
        },
        \\
        \bar{q}
        &=
        \frac{
        \mathbb{E}_{x}\Big\{
        \cosh^n\big(
        \beta \bar{m} + x\sqrt{\beta\alpha\bar{p}}
        \big)
        \tanh^2\big(
        \beta \bar{m} + x\sqrt{\beta\alpha\bar{p}}
        \big)
        \Big\}
        }{
        \mathbb{E}_{x}\Big\{
        \cosh^n\big(
        \beta \bar{m} + x\sqrt{\beta\alpha\bar{p}}
        \big)
        \Big\}
        },
        \\
        \bar{p}
        &=
        \frac{\beta\bar{q}}
        {\big[1-\beta(1-\bar{q})\big]\big[1-\beta(1-\bar{q})-n\beta\bar{q}\big]}.
    \end{aligned}
\end{equation}

This is exactly the expression previously obtained in \cite{dotsenko1994partial} by replica methods in the study of partial annealing. Here, however, it is recovered within an interpolation scheme and for $n$ treated directly as a real parameter, rather than as an integer to be continued analytically.

\begin{figure}[!t]
    \centering
    \includegraphics[width=15cm]{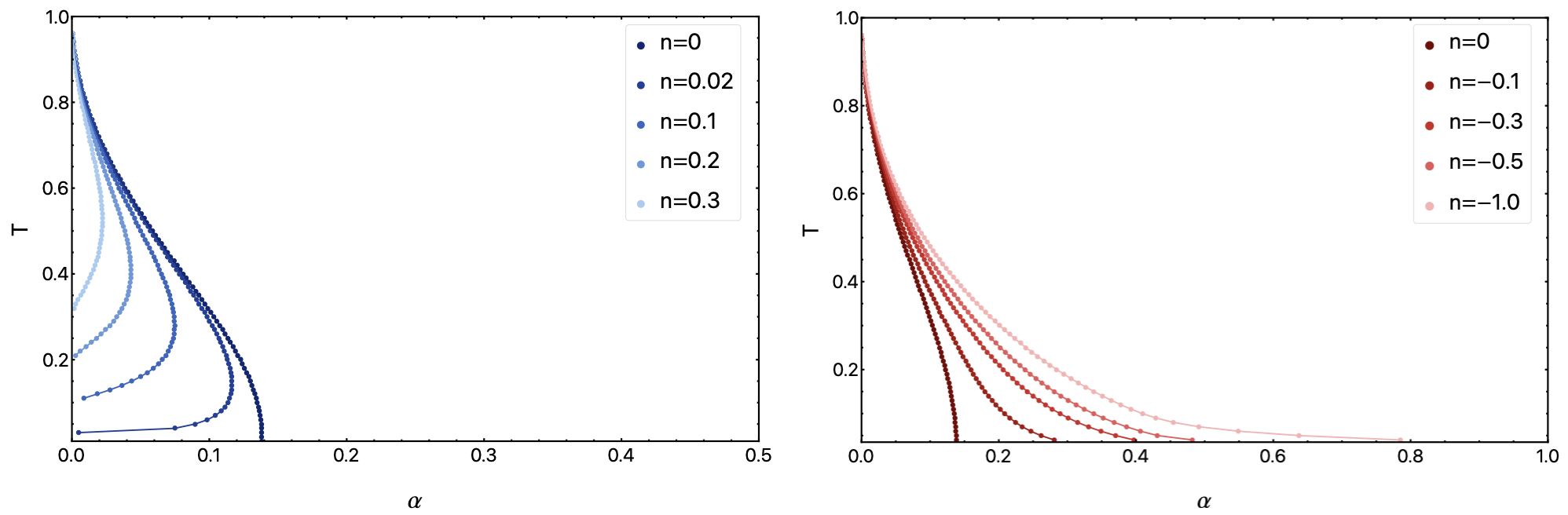}
    \caption{Retrieved critical line for different values of $n$ in the $(\alpha,T)$ plane. The case $n=0$ reproduces the standard quenched Hopfield model. As $n$ decreases and becomes more negative (right panel), the critical line moves toward larger values of the load $\alpha$, showing an enlargement of the retrieval region. Conversely, for positive values of $n$ (left panel), the retrieval region undergoes shrinkage, while RS instability is exacerbated at small values of $T$. This behavior is presumably attributable to the forced coalescence of patterns for $n>0$. Moreover, for $T<n$ the analytical results is not valid any more.}
    \label{fig:phase_diagram}
\end{figure}

\par\medskip
In Fig.~\ref{fig:phase_diagram} we report the numerical solution of \eqref{eq:self_n} for different non-positive values of $n$. The case $n=0$ corresponds to the standard Hopfield model and provides a benchmark for the consistency of our approach. We observe that, as $n$ becomes increasingly negative, the ferromagnetic region expands, eventually reaching $\alpha=1$ at zero temperature. Instead, for small but positive values of $n$, the retrieval region shrinks and, in particular, the numerical instability associated with the RS assumption becomes more pronounced.

\begin{figure}[t]
    \centering
    \includegraphics[width=15cm]{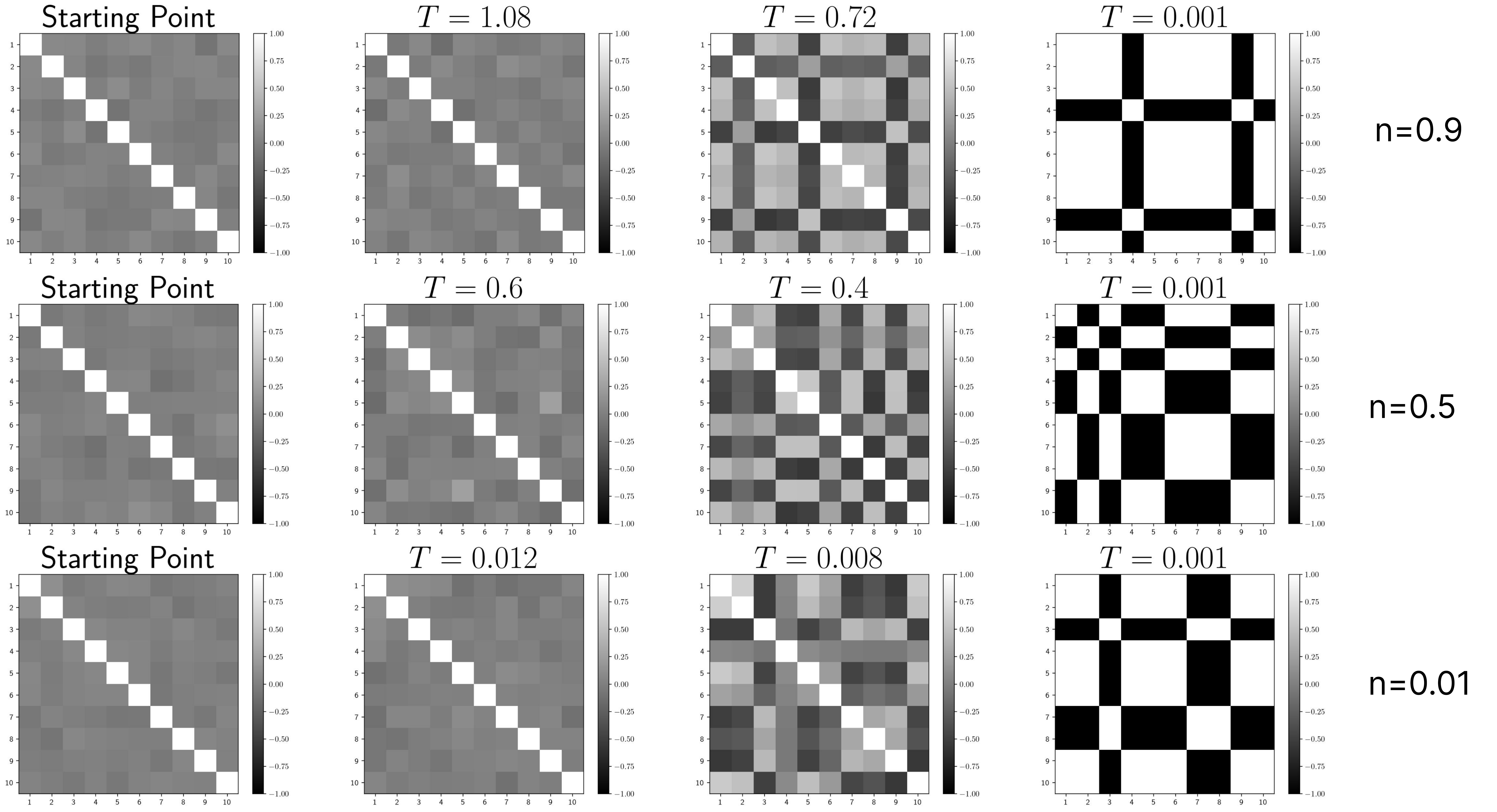}
    \caption{Evolution of the pattern-pattern correlation matrix for different values of the partial-annealing parameter $n$. Each row corresponds to a fixed value of $n$, while the columns show the initial condition and the configurations reached as the pattern temperature $T$ is decreased. For $T>n$, the off-diagonal correlations remain weak and the patterns can still be regarded as effectively uncorrelated. Conversely, when $T\leq n$, sizeable correlations emerge among the patterns, eventually leading, at very low temperature, to highly structured correlation patterns. This illustrates the breakdown of the independence assumption underlying the analytical computation outside the regime $T>n$.}
    \label{fig_corr}
\end{figure}

\begin{remark}
If we restrict to the case $n>0$, in order for the RS free energy \eqref{eq:free_n} and the corresponding self-consistency equations \eqref{eq:self_n} to be well defined, one must require that the arguments of the logarithms in \eqref{eq:free_n} are positive. This yields the conditions
\begin{equation}
    1-\beta(1-\bar q)>0,
    \qquad
    1-\beta(1-\bar q)-n\beta\bar q>0.
\end{equation}
Since $n>0$ and $\bar q\ge 0$, the second inequality is stronger and therefore implies the first one. Hence the relevant constraint is
\begin{equation}
    1-\beta\bigl(1-(1-n)\bar q\bigr)>0,
\end{equation}
or equivalently
\begin{equation}
    T>1-(1-n)\bar q.
\end{equation}
Moreover, because $0\le \bar q\le 1$, one has
\begin{equation}
    1-(1-n)\bar q \ge n,
\end{equation}
so that a necessary condition is
\begin{equation}
    T>n.
\end{equation}

This restriction is not merely technical, but identifies the regime in which the pattern variables can still be regarded as effectively independent for $n>0$. In this region, namely for $T>n$, partial annealing induces only a weak onset of correlations among the patterns, which remain sufficiently small, see Fig.~\ref{fig_corr}, thus supporting the assumptions used in the derivation.

By contrast, for $n>0$ and $0\leq T\leq n$, the pattern dynamics becomes genuinely relevant and, as shown in the next section, the patterns evolve towards a correlated state. In this regime, the independence assumptions underlying the RS computation are no longer justified. Therefore, for positive replica number, the analytical results obtained in this section should be interpreted as valid only in the regime $T>n$.

For $n\leq 0$, the situation is different: even though the patterns remain dynamical variables, their evolution is towards progressively more decorrelated configurations rather than towards an ordered correlated state. As a consequence, the assumptions underlying the RS computation are not spoiled by the pattern dynamics in that sector, and the RS results derived here remain consistent also for $n\leq 0$.
\end{remark}

\begin{remark}
The first limiting case to consider is the fully annealed one, corresponding
to $n=1$. Evaluating the RS free energy in this limit gives
\begin{equation}
\label{eq:free_n_1}
    \begin{aligned}
        f_1(\beta, \alpha)
        &=
        -\frac{1}{\beta}\log 2
        +\frac{\alpha}{2\beta}\log(1-\beta)
        -\frac{1}{\beta}\log\!\big(\cosh(\beta\bar m)\big)
        +\frac{1}{2}\bar m^2 .
    \end{aligned}
\end{equation}
Extremization with respect to the remaining order parameter yields
\begin{equation}
\label{eq:self_n__1_only_m}
    \begin{aligned}
        \bar{m}
        &=\tanh\big(\beta \bar{m}\big).
    \end{aligned}
\end{equation}
We recall that this expression is valid under the condition $T>n=1$, or
equivalently $\beta<1$. In this region, the only solution of
\eqref{eq:self_n__1_only_m} is $\bar m=0$. Therefore, the free energy reduces to
\begin{equation}
\label{eq:free_annealing}
    \begin{aligned}
        f_1(\beta, \alpha)
        &=
        -\frac{1}{\beta}\log 2
        +\frac{\alpha}{2\beta}\log(1-\beta),
    \end{aligned}
\end{equation}
which coincides with the well-known annealed free energy of the Hopfield
model~\cite{barra2012glassy, barra2008ergodic}. Consistently with the standard
annealed computation, the expression obtained through the partial-annealing
framework is therefore valid only in the high-temperature region $\beta<1$.
\end{remark}

\begin{remark}
The second limiting case is the quenched regime, obtained by taking
$n\to 0$. As anticipated above, this limit corresponds to fixed $T$ and
$T'\to\infty$ or, equivalently, $\beta'\to 0$. In this regime the pattern
dynamics becomes insensitive to the neuronal thermodynamics, and the standard
quenched Hopfield model is recovered. Indeed, starting from
\eqref{eq:free_n} and performing the limit $n\to 0$, one obtains
\begin{equation}
\label{eq:free_n0}
\begin{aligned}
     f_{0, \alpha}(\beta)
     &=
     -\frac{1}{\beta}\log 2
     -\frac{1}{\beta }\log\mathbb{E}_{x}\cosh\Big(
     \beta \bar m + x\sqrt{\beta \alpha \bar p}
     \Big)
     +\frac{\alpha}{2\beta}\log\Big(1-\beta(1-\bar q)\Big)
     \\
     &\qquad
     -\frac{\alpha}{2}\dfrac{\bar q}{1-\beta(1-\bar q)}
     +\frac{1}{2}\bar m^2
     +\frac{\alpha}{2}\bar p(1-\bar q).
\end{aligned}
\end{equation}
The corresponding order parameters are determined by the self-consistency
equations
\begin{equation}
\label{eq:self_n0}
    \begin{aligned}
        \bar{m}
        &=\mathbb{E}_{x}\Big\{
        \tanh\big(
        \beta \bar{m} + x\sqrt{\beta\alpha\bar{p}}
        \big)
        \Big\},
        \\
        \bar{q}
        &=\mathbb{E}_{x}\Big\{
        \tanh^2\big(
        \beta \bar{m} + x\sqrt{\beta\alpha\bar{p}}
        \big)
        \Big\},
        \\
        \bar{p}
        &=
        \frac{\beta\bar{q}}
        {\big[1-\beta(1-\bar{q})\big]^2}.
    \end{aligned}
\end{equation}
These expressions coincide with the replica-symmetric free energy and saddle
point equations of the standard quenched Hopfield model.
\end{remark}

\section{Mean-field MCMC dynamics for the slow pattern evolution}
\label{sec:numerical}

The analytical results of Sec.~\ref{sec:RS} characterize the thermodynamic effects of partial annealing and clarify the interplay between pattern and neuronal dynamics. In order to complement this picture, we now introduce a numerical dynamics aimed at simulating directly the evolution of the patterns $\bm\xi$ under the effective energy induced by the fast neuronal subsystem.

As discussed above, once the patterns are treated as slow dynamical variables, their effective energy is given by the free energy of the Hopfield model at fixed $\bm\xi$, namely
\begin{equation}
    \mathcal{E}_{N,P}(\bm\xi)
    =
    f_{N,P}(\beta\vert\bm\xi)
    =
    -\frac{1}{\beta}\log Z_N(\beta\vert\bm\xi).
\end{equation}
In principle, this suggests a Glauber-type Markov chain on the pattern space, based on single-entry flips of the variables $\xi_i^\mu$. More precisely, if $\tilde{\bm\xi}$ denotes the configuration obtained from $\bm\xi$ by flipping the component $\xi_i^\mu\mapsto-\xi_i^\mu$, then the corresponding energy variation is
\begin{equation}
    \Delta\mathcal{E}
    =
    \mathcal{E}_{N,P}(\tilde{\bm\xi})-\mathcal{E}_{N,P}(\bm\xi)
    =
    -\frac{1}{\beta}\Big(\log Z_N(\beta\vert\tilde{\bm\xi})-\log Z_N(\beta\vert\bm\xi)\Big)
    =
    -\frac{1}{\beta}\,\Delta\log Z_N.
\end{equation}
The associated Glauber acceptance probability is therefore
\begin{equation}
    \mathbb{P}(\xi_i^\mu\to-\xi_i^\mu)
    =
    \frac{1}{1+\exp\big(\beta'\Delta\mathcal{E}\big)}
    =
    \frac{1}{1+\exp\left(-\frac{\beta'}{\beta}\Delta\log Z_N\right)}
    =
    \frac{1}{1+\exp\left(-n\,\Delta\log Z_N\right)},
\end{equation}
where, as before, $n=\beta'/\beta$.

At this level, however, a direct implementation is computationally prohibitive. Indeed, for every attempted flip of a single pattern entry, one should evaluate the difference of two free energies, each requiring the computation of the partition function
\begin{equation}
    Z_N(\beta\vert\bm\xi)
    =
    \sum_{\bm\sigma\in\Omega}
    \exp\big(-\beta H_N(\bm\sigma\vert\bm\xi)\big),
\end{equation}
which involves a sum over all $2^N$ neural configurations. This makes an exact Glauber dynamics on $\bm\xi$ infeasible already for moderate system sizes.

For this reason, we introduce a mean-field approximation of the pattern dynamics, exploiting the fully-connected structure of the Hopfield model. The key numerical task is to estimate efficiently the increment $\Delta\log Z_N$ produced by a single flip of $\xi_i^\mu$. To this end, we use a first-order expansion:
\begin{equation}
    \Delta\log Z_N
    \approx
    \frac{\partial \log Z_N}{\partial \xi_i^\mu}\,\Delta\xi_i^\mu.
\end{equation}
Since $\Delta\xi_i^\mu=-2\xi_i^\mu$, one finds
\begin{equation}
    \Delta\log Z_N
    \approx
    \beta \omega_{N, \bm \xi} (m_\mu \sigma_i)\,\Delta\xi_i^\mu
    =
    -2\beta\,\xi_i^\mu \omega_{N,\bm\xi}(m_\mu \sigma_i),
\end{equation}
where the thermal average at fixed $\bm\xi$ is defined by
\begin{equation}
    \omega_{N,\bm\xi}\big(f(\bm\sigma,\bm\xi)\big)
    =
    \frac{\sum_{\bm\sigma\in\Omega}f(\bm\sigma,\bm\xi)e^{-\beta H_N(\bm\sigma\vert\bm\xi)}}
    {Z_N(\beta\vert\bm\xi)}.
\end{equation}
From now on, we imply the dependence on $\bm \xi$ and $N$ from $\omega_{N, \bm \xi}(\cdot)$.

In order to close the dynamics, we further exploit the mean-field character of the model and approximate the mixed correlation through the factorized ansatz
\begin{equation}
    \omega(m_\mu \sigma_i)
    \approx
    \bar m_\mu\,\omega(\sigma_i).
\end{equation}
Introducing the local field
\begin{equation}
    h_i
    =
    \sum_{\mu=1}^P \bar m_\mu \xi_i^\mu,
\end{equation}
we use the standard mean-field relations
\begin{equation}
    \omega(\sigma_i)
    =
    \tanh(\beta h_i),
    \qquad
    \bar m_\mu
    =
    \frac{1}{N}\sum_{i=1}^N \xi_i^\mu \omega(\sigma_i).
\end{equation}
Combining these approximations, the Glauber transition probability becomes
\begin{equation}
\mathbb{P}(\xi_i^\mu\to-\xi_i^\mu)
=
\frac{1}{1+\exp\left(2\beta n\,\xi_i^\mu \bar m_\mu \omega(\sigma_i)\right)}.
\label{eq:agg_en_xi}
\end{equation}

This rule has the expected limiting behavior. In particular, for $n=0$ one obtains
\begin{align}
\mathbb{P}(\xi_i^\mu\to-\xi_i^\mu)=\frac{1}{2},
\end{align}
namely unbiased random flips of the pattern variables. This is fully consistent with the interpretation of $n=0$ as the quenched limit $\beta'\to 0$, corresponding to an infinite temperature in the slow sector.

The resulting algorithm (sketched in Algorithm \ref{alg:mf_mcmc_patterns}) should therefore be understood as an effective mean field MCMC dynamics for the slow variables: it does not simulate the exact Glauber chain associated with the free energy $\mathcal{E}_{N,P}(\bm\xi)$, but rather a tractable approximation designed to capture its qualitative behavior, and in particular the reorganization of the patterns induced by negative values of $n$.

\begin{algorithm}[h]
\caption{Mean-field MCMC dynamics for the pattern variables}
\label{alg:mf_mcmc_patterns}
\begin{algorithmic}[1]
\Require $N$, $P$, $\beta$, $n$, number of sweeps $T_{\max}$
\Ensure Evolved pattern configuration $\bm\xi$

\State Initialize $\bm\xi\in\{-1,+1\}^{N\times P}$ 
\State Initialize $\bm\sigma^{(0)}$ close to one of the selected patterns
\State Compute initial magnetizations $m_\mu^{(0)}$, for $\mu=1,\dots,P$

\For{$t=1$ to $T_{\max}$}
    \State Compute local fields
    \[
    h_i=\sum_{\mu=1}^P \bar m_\mu\,\xi_i^\mu,
    \qquad i=1,\dots,N
    \]
    \State Compute mean value of spins
    \[
    \omega(\sigma_i)=\tanh(\beta h_i),
    \qquad i=1,\dots,N
    \]
    \State Update mean-field magnetizations
    \[
    \bar m_\mu=\frac{1}{N}\sum_{i=1}^N \xi_i^\mu \omega(\sigma_i),
    \qquad \mu=1,\dots,P
    \]

    \For{each pair $(i,\mu)$, or for a random subset of pairs}
        \State Estimate the free-energy increment
        \[
        \Delta\log Z_N \approx -2\beta\,\xi_i^\mu \bar m_\mu \omega(\sigma_i)
        \]
        \State Compute the Glauber flip probability
        \[
        p_{i\mu}
        =
        \frac{1}{1+\exp\left(2\beta n\,\xi_i^\mu \bar m_\mu \omega(\sigma_i)\right)}
        \]
        \State Draw $u\sim \mathrm{Uniform}(0,1)$
        \If{$u<p_{i\mu}$}
            \State Flip the pattern entry:
            \[
            \xi_i^\mu \leftarrow -\xi_i^\mu
            \]
        \EndIf
    \EndFor

    \State Check convergence criterion or stop if the maximum number of sweeps is reached
\EndFor

\State \Return $\bm\xi$
\end{algorithmic}
\end{algorithm}

In the following subsection we test numerically the prediction that negative values of $n$ induce an effective orthogonalization of the stored patterns. 

\subsection{Numerical evidence for pattern orthogonalization under partial annealing}

We now present three numerical experiments aimed at testing the physical picture suggested by the analytical results and by the original analysis of \cite{dotsenko1994partial}, namely that negative values of the parameter $n$ favor a progressive decorrelation of the stored patterns. More precisely, our goal is to understand whether the slow mean field MCMC dynamics introduced above effectively reshapes the pattern ensemble so as to reduce interference, and whether this geometric rearrangement translates into an improved retrieval performance.

The results reported below address this question from three complementary viewpoints. First, we directly monitor the evolution of the pattern-pattern correlation matrix and show that the slow dynamics behaves qualitatively differently for positive and negative values of $n$. Second, we investigate how this rearrangement affects the retrieval process in the presence of strongly biased patterns, where the standard Hopfield dynamics is known to fail. Finally, we compare the partial annealing mechanism with a classical decorrelation strategy from the neural network Literature, namely the pseudoinverse prescription, showing that the former yields larger and more stable attraction basins in the highly biased regime.

\paragraph{Direct observation of pattern decorrelation}

Our first experiment is designed to probe directly the effect of the slow dynamics on the mutual correlations among the stored patterns. To this end, we ran the mean field MCMC algorithm for two representative values of the control parameter, namely $n=+1$ and $n=-1$, and compared the pattern configurations at the beginning and at the end of the dynamics.

In the case $n<0$, we initialized the system from a strongly correlated configuration, choosing in particular the first two patterns so that
\[
\frac{1}{N}\sum_{i=1}^N \xi_i^1\xi_i^2 \approx 0.85.
\]
We then evolved the slow dynamics until thermalization and measured the pattern-pattern correlation matrix both at the initial and at the final time. The corresponding heatmaps show a clear decorrelation effect: the initially correlated patterns progressively lose their mutual overlap, and the final pattern ensemble becomes significantly closer to an orthogonal configuration.

By contrast, for $n>0$ we started from a standard Rademacher initialization, namely from an essentially uncorrelated pattern set. In this case, the final heatmaps reveal the opposite trend: the slow dynamics tends to induce positive correlations among the patterns, indicating that the effective interaction in the pattern sector is no longer repulsive, but instead favors clustering or alignment.

\begin{figure}[t]
    \centering
    \includegraphics[width=12cm]{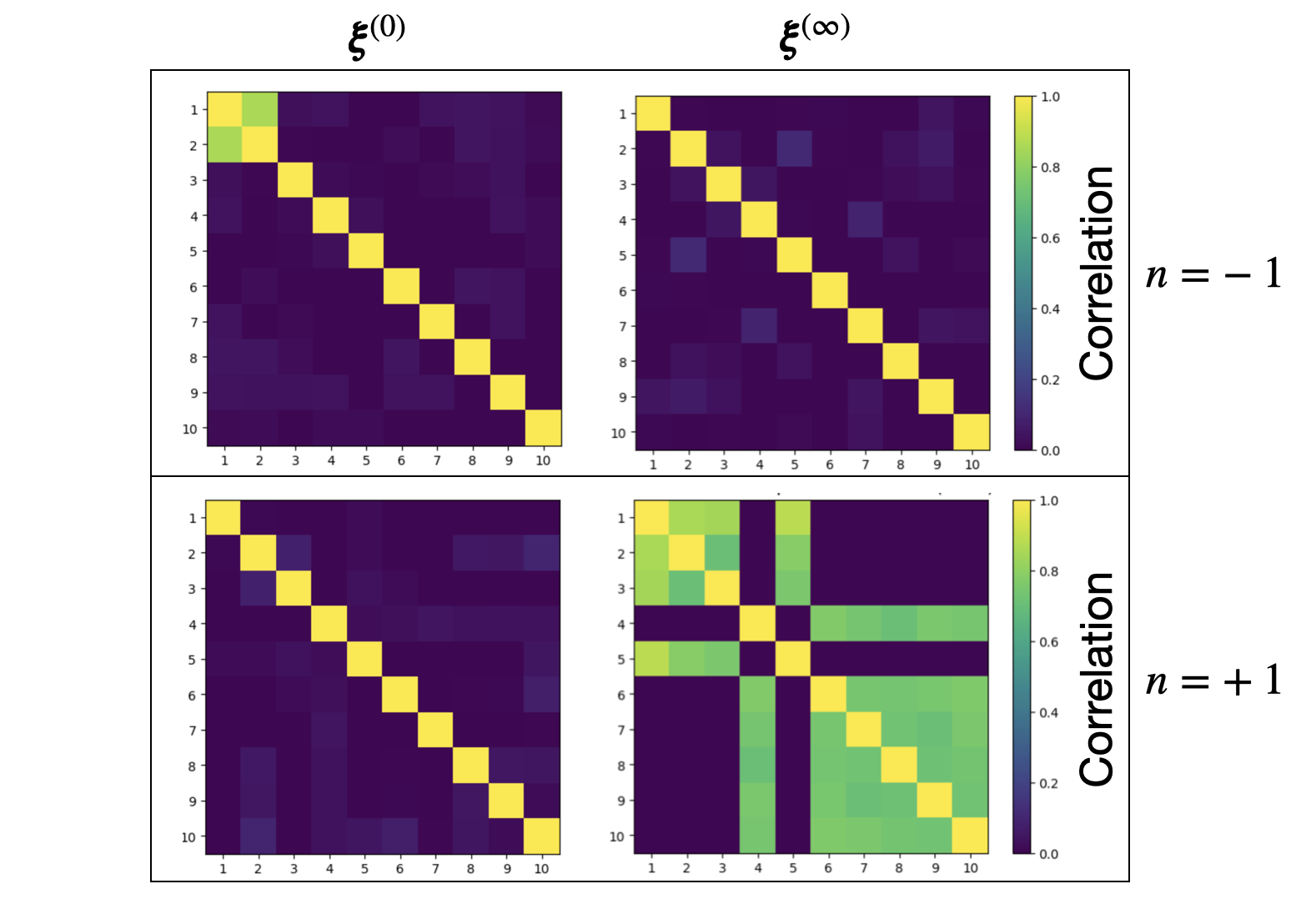}
    \caption{Pattern-pattern correlation matrices at the initial and final stages of the slow dynamics. For $n=-1$, the initially first and second correlated patterns progressively decorrelate, and the final configuration is closer to an orthogonal pattern ensemble. For $n=+1$, starting from a standard Rademacher initialization, the opposite effect is observed: the dynamics tends to build positive correlations among the patterns. For both experiment we use $N=500$, $K=100$, $\beta=1000$.}
    \label{fig:evol}
\end{figure}

These results, presented in Fig.~\ref{fig:evol}, provide a direct geometric visualization of the role of the parameter $n$. Negative values of $n$ drive the pattern ensemble toward a more orthogonal arrangement, while positive values favor the buildup of correlations. In this sense, the numerical dynamics supports the interpretation of the $n<0$ regime as an effective unlearning or anti-Hebbian mechanism acting on the slow degrees of freedom.

\paragraph{Recovery of retrieval in the presence of biased patterns}

The second experiment addresses the functional consequence of this decorrelation mechanism. To make the effect more transparent, we considered an ensemble of initially biased patterns, sampled according to
\begin{equation}
\mathbb{P}\Big(\xi_i^{\mu, (0)}\Big)
=
\frac{1+b}{2}\delta\Big((\xi_i^{\mu, (0)}-1\Big)
+
\frac{1-b}{2}\delta\Big((\xi_i^{\mu, (0)}+1\Big),
\end{equation}
with $b>0$ quantifying the degree of bias.

This choice produces a highly non-ideal storage set: the patterns are no longer centered, share a common positive magnetization, and therefore generate a strong spurious collective direction. As a consequence, if one directly applies the standard Hopfield retrieval dynamics to the initial pattern set, retrieval fails. Rather than being attracted to a single stored pattern, the neuronal configuration is pulled toward the average activity of the pattern ensemble. In particular, one observes that the Mattis magnetizations do not select a unique retrieved state, but instead remain of the order
\[
m_\mu \approx b^2,
\qquad \forall \mu,
\]
which reflects the global bias of the dataset rather than genuine associative recall.

We then applied the partial annealing dynamics with $n<0$ to the same initial pattern set, and subsequently ran a standard Monte Carlo (MC) retrieval dynamics on the final evolved patterns. In this case, the phenomenology changes drastically. After the slow rearrangement of the pattern ensemble, the retrieval dynamics is restored: the neuronal state aligns with the nearest stored pattern, and the corresponding Mattis magnetization approaches one, while the overlaps with the other patterns remain small.

\begin{figure}[t]
    \centering
    \includegraphics[width=13cm]{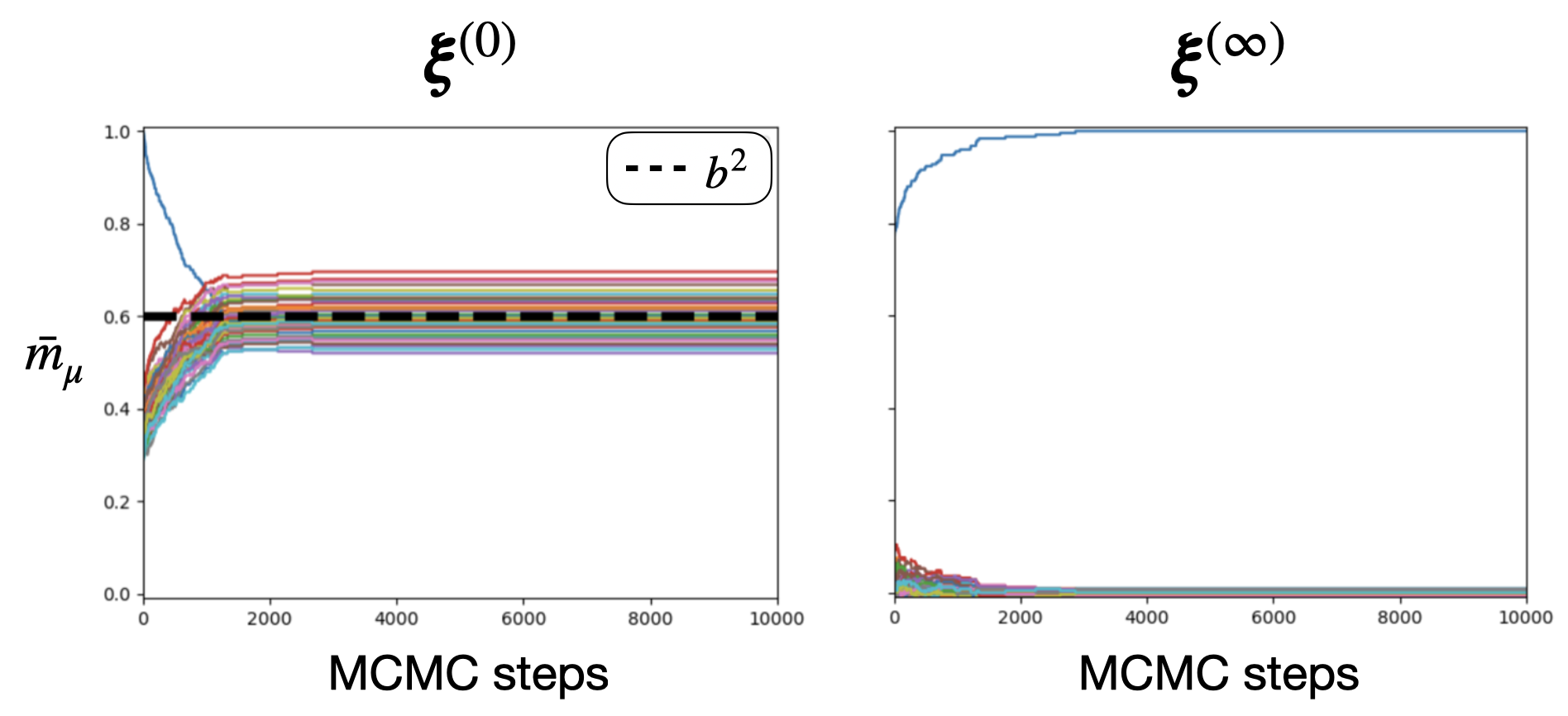}
    \caption{Retrieval dynamics in the presence of biased patterns with $b=0.8$. Left: for the initial biased pattern set, the neuronal configuration is not attracted to a single stored pattern, but rather to the average activation direction generated by the bias. Right: after partial annealing with $n<0$, standard retrieval is restored, and the neuronal state aligns with the closest stored pattern.}
    \label{fig:biased_pattern}
\end{figure}

Fig.~\ref{fig:biased_pattern} shows that the main effect of partial annealing is not merely geometric. By reducing the mutual interference induced by the common bias, the slow dynamics reshapes the stored memories into a form that is again compatible with reliable associative recall. In particular, the numerical evidence indicates that the orthogonalization tendency observed for $n<0$ has a direct dynamical counterpart at the level of retrieval.

\paragraph{Comparison with the pseudoinverse prescription}

The third experiment places the partial-annealing mechanism in the broader context of decorrelation and unlearning procedures for associative memories. In particular, we compare it with the well-known pseudoinverse prescription \cite{kanter1987associative, personnaz1986biologically}, in which the Hebbian coupling matrix is replaced by
\begin{equation}
J_{ij}^{\mathrm{KS}}
=
\frac{1}{N}\sum_{\mu,\nu=1}^{P}
(\xi_i^\mu-b)\,(C^{-1})_{\mu\nu}\,(\xi_j^\nu-b),
\end{equation}
where
\begin{equation}
C_{\mu\nu}
=
\frac{1}{N}\sum_{i=1}^N (\xi_i^\mu-b) (\xi_i^\nu-b)
\end{equation}
is the pattern correlation matrix.
This construction is a classical way to compensate for pattern correlations and to enlarge the retrieval region by explicitly inverting the overlap structure of the stored set \cite{agliari2024regularization, fachechi2024outperforming, agliari2019dreaming, fachechi2019dreaming}.

In the biased regime described above, we compared the retrieval properties of the pseudoinverse model with those of the pattern ensemble produced by partial annealing with $n<0$. The comparison was performed numerically by reconstructing the corresponding retrieval maps and estimating the associated attraction basins.

\begin{figure}[t]
    \centering
    \includegraphics[width=15.3cm]{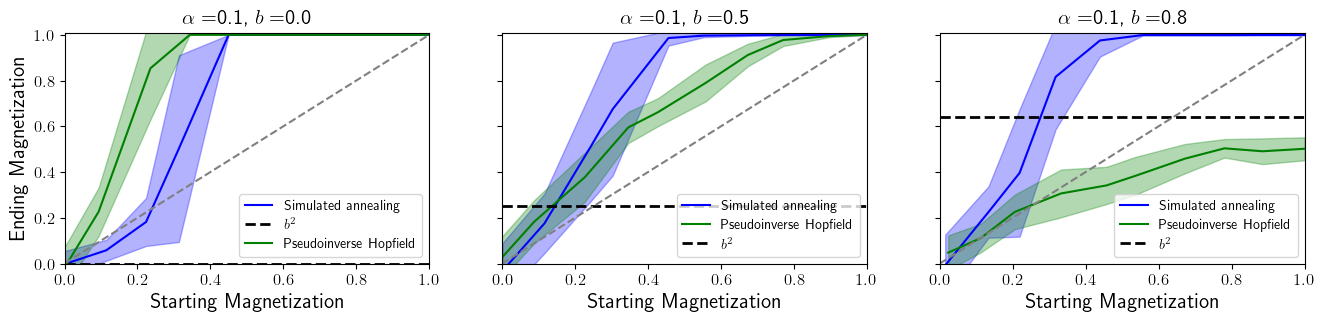}
    \caption{Comparison between partial annealing and the pseudoinverse prescription in the biased regime. The retrieval maps show that, for sufficiently large bias, the attraction basins generated by partial annealing are larger and more stable than those obtained with the pseudoinverse coupling matrix.}
    \label{fig:pseudo_inv}
\end{figure}

The outcome (see Fig.~\ref{fig:pseudo_inv}) is particularly significant in the strongly biased regime. While both mechanisms improve retrieval with respect to the naive Hebbian prescription, the partial-annealing dynamics systematically produces larger and more stable attraction basins than the pseudoinverse model. In other words, the decorrelation induced dynamically in the slow sector appears to be more effective, at high bias, than the static correction implemented through the inverse correlation matrix.

This result suggests that partial annealing is not simply another decorrelation recipe, but rather a genuinely adaptive mechanism: instead of correcting correlations a posteriori at the level of the couplings, it progressively reshapes the pattern ensemble itself under the effective free-energy landscape generated by the neuronal subsystem. The resulting memory structure is therefore both less frustrated and more robust under retrieval dynamics.

\par\medskip
Taken together, these three experiments support a coherent picture. Negative values of $n$ induce a slow reorganization of the stored patterns toward less correlated configurations; this decorrelation suppresses the spurious collective directions generated by bias; and the corresponding reduction in interference translates into a substantial enhancement of the retrieval capabilities of the network. In particular, the numerical results strongly support the interpretation of the $n<0$ regime as an effective orthogonalization mechanism for the stored memories.

\section{Conclusions and outlooks}
\label{sec:conclusions}
In this work, we have developed a theoretical framework for analysing partially annealed associative neural networks. This framework is based on Guerra’s interpolation scheme and is formulated directly for a real replica parameter $n$. By avoiding the need for analytic continuation from integer replicas, our approach provides a transparent characterisation of the RS pseudo-replicated free energy. 

\par\medskip
From a physical perspective, the parameter $n$, interpreted as the ratio between the temperatures of the slow and fast noise, emerges as a key control variable governing the interplay between neuronal dynamics and pattern evolution. In particular, once checked that the limits of $n\to 1$ and $n\to0$ return the annealed and quenched pictures, the regime $n < 0$, corresponding to an effective negative temperature in the slow noise, has been shown to induce a non-trivial reorganisation of the stored patterns \cite{dotsenko1994partial}.

Further corroborated  by numerical simulations, the findings of this study demonstrate that negative values of n induce a progressive decorrelation of the pattern ensemble. This phenomenon can be interpreted as an effective orthogonalisation mechanism acting on the slow degrees of freedom, thereby reducing interference among patterns and reshaping the underlying energy landscape. Consequently, the retrieval region is expanded, and the system displays enhanced robustness against high storage loads and pattern bias.

These findings suggest that partial annealing should be regarded not simply as a technical generalisation of quenched and annealed limits, but as a genuinely adaptive mechanism capable of reorganising memory representations in a self consistent manner achieving maxima storage capacity.

\par\medskip
Several avenues for future research naturally arise from this study. On the theoretical side, it would be particularly valuable to analyse the stability of the replica symmetric solution against one-step replica symmetry breaking, for instance by deriving the corresponding critical line in the spirit of De Almeida and Thouless \cite{de1978stability}.

On the dynamical side, a more refined treatment of the slow evolution, beyond the mean-field approximation, could shed further light on the microscopic mechanisms underlying pattern orthogonalisation. Finally, the connections with modern machine learning systems, where parameters evolve on multiple timescales, may deserve deeper exploration, particularly in relation to regularisation and representation learning.

\ack{L.A. acknowledges funding from the project “Patto Territoriale del Sistema Universitario Pugliese” (CUP F61B23000370006). \\
A.B. acknowledges funding from Sapienza University: projects prot. n. RM12419112BF7119 and prot. n.  RM12519999AB8CA9 \\
A.A. acknowledges BULBUL “Brain-inspired ULtra-Fast \& Ultra-sharp neural networks” for support via post-Lauream research fellowships 
“Statistical mechanics of hetero-associative neural networks” (CUP F85F21006230001, D.D. n. 325/29-09-2025). \\
L.A., A.A., A.B., S.F. are members of the GNFM group within INdAM which is acknowledged too.}



\data{No data are associated with this article.}


\appendix 
\section{Computations of the $t-$derivatives and the one body terms}
\label{app:tderiv}

Let us start by the computation of the derivative of the replicated free energy with respect to $t$. 

Starting from the derivation of the expression \eqref{eq:dtf_0}, namely
\begin{equation}
\begin{array}{lll}
    d_t f_{n,N}(\beta; t) &=& 
    \dfrac{1}{  N}\mathbb{E}\left\{ \dfrac{Z_N^n(\beta; t \vert \bm \xi)}{\mathbb{E} Z_N^n(\beta; t \vert \bm \xi)}\dfrac{1}{Z_N(\beta; t \vert \bm \xi)}\SOMMA{\{\bm\sigma\}}{}e^{-\beta H_N(\bm\sigma, \bm z;t|\bm\xi)}\partial_t H_N(\bm\sigma, \bm z;t|\bm\xi)\right\}
    \\\\
    &=& 
    \dfrac{1}{N}\mathbb{E}\Bigg\{ W_{n,N}(\beta;t|\bm\xi)\omega\Big(\partial_t H_N(\bm\sigma, \bm z;t|\bm\xi)\Big)\Bigg\},
     \label{eq:dtf_0_app}
\end{array}
\end{equation}
where one can find the expression of $\partial_t H_N$ in \eqref{eq:dtf_1}, we can recall the definition of the quenched average as $\l\cdot\r = \mathbb{E}[W_{n,N}\omega_t(\cdot)]$ in order to get
\begin{equation}
\begin{array}{lll}
   d_t f_{n,N}(\beta; t) &=& - \dfrac{1}{2}\l (m_1)^2\r +\dfrac{\psi}{\beta} \l m_1 \r -\dfrac{1}{2N}\sqrt{\dfrac{1}{t N\beta}} \SOMMA{\mu,i}{}\mathbb{E}[\xi_i^\mu W_{n,N} \omega(\sigma_i z_\mu)]
    \\\\
    && \hspace{-1cm} +\dfrac{ B}{2N\beta\sqrt{1-t}}\SOMMA{\mu}{} \mathbb{E}[Y_{\mu} W_{n,N}\omega(z_{\mu})]+\dfrac{A}{2N\beta\sqrt{1-t} }\SOMMA{i}{}\mathbb{E}[J_{i} W_{n,N} \omega( \sigma_{i})]+\dfrac{1}{2N\beta} C\SOMMA{\mu}{} \l z_{\mu}^2 \r.
     \label{eq:dtf_1.2}
\end{array}
\end{equation}
Now we can exploit the relation
\begin{align}
\mathbb{E}_{\bm J}( \bm J f(\bm J) ) = \mathbb{E}_{\bm J} \Bigg( \frac{\partial}{\partial \bm J} f(\bm J) \Bigg),
\label{eq:steins_app}
\end{align}
valid for any smooth function $f(\bm J)$ of a centered unit Gaussian variable $\bm J$ for which the two expectations $\mathbb{E}\left( \bm J f(\bm J)\right)$ and $\mathbb{E}\left( \partial_{\bm J} f(\bm J)\right)$ both exist.

In this way we find
\begin{align}
    \mathbb{E}[\xi_i^\mu W_{n,N} \omega(\sigma_i z_\mu)] &= \mathbb{E}[\partial_{\xi_i^\mu}\left(W_{n,N} \omega(\sigma_i z_\mu)\right)] = \mathbb{E}[\partial_{\xi_i^\mu}\left(W_{n,N} \omega(\sigma_i z_\mu)\right)] \notag \\
    &=  \mathbb{E}[\partial_{\xi_i^\mu}\left(W_{n,N} \right)\omega(\sigma_i z_\mu) +W_{n,N} \partial_{\xi_i^\mu}\left( \omega(\sigma_i z_\mu) \right)]; \\
    \partial_{\xi_i^\mu}\left(W_{n,N} \right) &= n W_{n, N} \omega^2(\sigma_i z_\mu)  \sqrt{\dfrac{t}{N\beta}} \\
    \partial_{\xi_i^\mu}\left( \omega(\sigma_i z_\mu) \right) &= \left[\omega(\sigma_i z_\mu)^2 - \omega^2(\sigma_i z_\mu)\right] \sqrt{\dfrac{t}{N\beta}}= \left[\omega(z_\mu^2)- \omega^2(\sigma_i z_\mu)\right]\sqrt{\dfrac{t}{N\beta}}; \\
    \mathbb{E}[\xi_i^\mu W_{n,N} \omega(\sigma_i z_\mu)] &=  \sqrt{\dfrac{t}{N\beta}}\mathbb{E}\left[ W_{n,N}\omega(z_\mu^2) +(n-1)W_{n,N} \omega^2(\sigma_i z_\mu) \right] \label{eq:E1_app}
\end{align}
and, analogously, 
\begin{align}
    \mathbb{E}[Y_{\mu} W_{n,N} \omega(z_{\mu})] &= \sqrt{B (1-t)} \mathbb{E}\left[ W_{n,N} \omega(z_\mu^2)+(n-1) W_{n,N} \omega^2(z_\mu)\right] \\
    \mathbb{E}[J_{i} W_{n,N} \omega( \sigma_{i})] &= \sqrt{A(1-t)}\mathbb{E}\left[ W_{n,N} \omega(\sigma_i^2)+(n-1) W_{n,N} \omega^2(\sigma_i)\right].
    \label{eq:E3_app}
\end{align}

Putting \eqref{eq:E1_app}-\eqref{eq:E3_app} together, we get
\begin{equation}
\begin{array}{lll}
   d_t f_{n,N}(\beta; t) &=& - \dfrac{1}{2}\Bigg[\l (m_1)^2\r -\dfrac{2\psi}{\beta} \l m_1 \r \Bigg]-\dfrac{K}{2N} [\l p_{11}\r- (1-n) \l q_{12}p_{12}\r]
    \\\\
    && +\dfrac{ B^2K}{2\beta N}[\l p_{11}\r-(1-n)\l p_{12}\r]+\dfrac{A^2}{2\beta}[1-(1-n)\l q_{12}\r]+\dfrac{K}{2\beta N} C \l p_{11}\r
     \label{eq:dtf_app}
\end{array}
\end{equation}
which is the same expression in \eqref{eq:dtf}.

We want to exploit the RS ansatz, namely the order parameters self-average around their mean value, for any order parameter $X$ whose mean value is $\bar X$, 
\begin{align}
    \lim_{N \to +\infty} \mathbb{P}_N(X)=\delta(X-\bar X)
\end{align}
and the variance vanishes in the thermodynamic limit
\begin{align}
    \lim_{N \to +\infty} \l (X-\bar X)^2 \r =0.
\end{align}
Exploiting some algebraic manipulation, one can say that, as $N \to +\infty$,
\begin{align}
\l (m_1-\mb)^2 \r \to 0 &\Rightarrow\l (m_1)^2\r - 2\bar{m}\l m_1\r = -\bar{m}^2 \\
\l (p_{12} - \bar p )(q_{12}-\qb)\r \to 0 &\Rightarrow \l q^{12} p^{12}\r - \bar{q}\l p^{12}\r  - \bar{p}\l q^{12}\r= -\bar{q}\bar{p}.
\end{align}
Now, if we set the constants as 
\begin{equation}
    \begin{array}{lll}
         \psi = \beta \bar{m}, && C = \beta(1-\bar{q})
         \\\\
         B ^2 = \beta \bar{q}, &&A^2 =  \alpha\beta \bar{p},
    \end{array}
\end{equation}
and if we recall the load of the network as $\alpha=\lim_{N \to +\infty} \dfrac{K}{N}$, 
Eq. \eqref{eq:dtf} becomes
\begin{equation}
\begin{array}{lll}
    d_t f_{n}(\beta; t) &=& \dfrac{1}{2}\bar{m}^2+\dfrac{1}{2} \alpha \bar{p}(1-\bar{q}) +\dfrac{\alpha}{2} n \bar{p}\bar{q}.
    \label{eq:dtffapp}
\end{array}
\end{equation}

Regarding the one body term, we compute it exploiting the fact that the two kind of variables are now independent from one other. This allows us to write the expression of the average of the one body term as 
\begin{equation}
\begin{array}{lll}
    &&\mathbb{E}Z_N^n(\beta, \bm \xi; t=0) = \mathbb{E} \Bigg\{\SOMMA{\{\bm\sigma\}}{}\displaystyle\int d\mu(z_\mu) \exp\Bigg[\psi N m_1 + B\SOMMA{\mu}{}Y_{\mu} z_{\mu}+ A\SOMMA{i}{}J_{i} \sigma_{i}+ \dfrac{C}{2} \SOMMA{\mu}{} z_{\mu}^2\Bigg]\Bigg\}^n
    \\\\
    &&= \mathbb{E} \Bigg\{\SOMMA{\{\bm\sigma\}}{}\exp\Bigg[\psi \SOMMA{i}{}\xi_i^1\sigma_i + A\SOMMA{i}{}J_{i} \sigma_{i}\Bigg]\displaystyle\int d\mu(z_\mu) \exp\Bigg[ \dfrac{C}{2} \SOMMA{\mu}{} z_{\mu}^2+ B\SOMMA{\mu}{}Y_{\mu} z_{\mu}\Bigg]\Bigg\}^n
    \\\\
    &&= \mathbb{E} \Bigg\{ \prod\limits_i\SOMMA{\{\sigma\}}{}\exp\Bigg[\psi \SOMMA{i}{}\xi_i^1\sigma_i + A\SOMMA{i}{}J_{i} \sigma_{i}\Bigg]\Bigg\}^n\mathbb{E} \Bigg\{\prod\limits_{\mu}\displaystyle\int \dfrac{d z_\mu}{\sqrt{2\pi}}\exp\Bigg[-\dfrac{1}{2}(1-C)  z_{\mu}^2+ B Y_{\mu} z_{\mu}\Bigg]\Bigg\}^n
    \\\\
    &&= 2^{nN}(1-C)^{-nK/2}\prod\limits_i\mathbb{E} \Bigg\{\cosh\Bigg[\psi \xi^1 + A J \Bigg]\Bigg\}^n \\\\
    &&\hspace{3cm}\cdot\prod\limits_{\mu>1}\mathbb{E}_{Y} \Bigg\{\displaystyle\int \dfrac{d x}{\sqrt{2\pi}}\exp\Bigg[-\dfrac{1}{2}  x^2+ \dfrac{B Y}{\sqrt{1-C}} x\Bigg]\Bigg\}^n
    \\\\
    &&= 2^{nN}(1-C)^{-nK/2}\prod\limits_i\mathbb{E} \Bigg\{\cosh\Bigg[\psi \xi^1 + A J \Bigg]\Bigg\}^n\prod\limits_{\mu>1}\mathbb{E}_{Y} \Bigg\{\exp\Bigg[ \dfrac{B^2 Y^2}{2(1-C)} \Bigg]\Bigg\}^n
    \\\\
    &&= 2^{nN}(1-C)^{-nK/2}\prod\limits_i\mathbb{E} \Bigg\{\cosh\Bigg[\psi \xi^1 + A J \Bigg]\Bigg\}^n\prod\limits_{\mu>1}\times\exp\Bigg\{\log\mathbb{E}_{Y} \exp\Bigg[ \dfrac{ n B^2 Y^2}{2(1-C)} \Bigg]\Bigg\}
    \\\\
    &&= 2^{nN}(1-C)^{-nK/2}\prod\limits_i\mathbb{E} \Bigg\{\cosh\Bigg[\psi \xi^1 + A J \Bigg]\Bigg\}^n\exp\Bigg\{ K\log\Bigg[ \frac{1}{\sqrt{1-\frac{n B^2}{1-C}}} \Bigg]\Bigg\}
\end{array}
\end{equation}
therefore we get
\begin{equation}
\begin{array}{lll}
   f_{n}(\beta; t=0) &=& -\dfrac{1}{\beta}\log 2-\dfrac{1}{\beta n}\log\mathbb{E}_{x}\cosh^n\Bigg[  \psi  + x\sqrt{A^2}\Bigg] 
   \\\\
   &&
   +\dfrac{\alpha}{2\beta }\log\Big(1-C\Big)-\dfrac{\alpha}{2\beta n}\log\Bigg(1-\dfrac{nB^2}{1-C}\Bigg).   
\end{array}
\label{eq:ff0app}
\end{equation}
which is the expression reported in \eqref{eq:one_body}.

\end{document}